# An Effective End-to-End Modeling Approach for Mispronunciation Detection

*Tien-Hong Lo, Shi-Yan Weng, Hsiu-Jui Chang, and Berlin Chen*

National Taiwan Normal University, Taiwan

{teinhonglo, 40547041S, 60647061S, berlin}@ntnu.edu.tw

## Abstract

Recently, end-to-end (E2E) automatic speech recognition (ASR) systems have garnered tremendous attention because of their great success and unified modeling paradigms in comparison to conventional hybrid DNN-HMM ASR systems. Despite the widespread adoption of E2E modeling frameworks on ASR, there still is a dearth of work on investigating the E2E frameworks for use in computer-assisted pronunciation learning (CAPT), particularly for mispronunciation detection (MD). In response, we first present a novel use of hybrid CTC-Attention approach to the MD task, taking advantage of the strengths of both CTC and the attention-based model meanwhile getting around the need for phone-level forced-alignment. Second, we perform input augmentation with text prompt information to make the resulting E2E model more tailored for the MD task. On the other hand, we adopt two MD decision methods so as to better cooperate with the proposed framework: 1) decision-making based on a recognition confidence measure or 2) simply based on speech recognition results. A series of Mandarin MD experiments demonstrate that our approach not only simplifies the processing pipeline of existing hybrid DNN-HMM systems but also brings about systematic and substantial performance improvements. Furthermore, input augmentation with text prompts seems to hold excellent promise for the E2E-based MD approach.

## 1. Introduction

The increasing development of automatic speech recognition (ASR) has led to its growing applications on computer-assisted pronunciation learning (CAPT). Paramount to the success of a CAPT system is the accuracy of the mispronunciation detection (MD) module, which manages to pinpoint erroneous pronunciations in the utterance of a second-language (L2) learner in response to a text prompt.

A common practice for mispronunciation detection is to extract decisive features (attributes) [1] from the prediction output of acoustic models which normally are estimated based on certain criteria that maximize the ASR performance. Although hidden Markov models with Gaussian mixture models accounting for state (or senone) emission probabilities (denoted by GMM-HMM) used to be the predominating approach for building the acoustic models involved in the mispronunciation detection process, a recent school of thought is to leverage various state-of-the-art deep neural network (DNN) architectures in place of GMM for modeling the state emission probabilities in HMM (denoted by hybrid DNN-HMM) [2-4], which shows good success in improving empirical performance [5-7]. As for decisive feature extraction, log-likelihood, log-posterior probability, segment duration-based score, and among others [8], are frequently used in evaluating phone- [9] or word-level [10] pronunciation quality, while log-posterior probability based scores and its prominent extension, namely goodness of pronunciation (GOP) [11, 12], are the most prevalent and have been shown to correlate well with human assessments. Yet there still is a wide array of studies that capitalize on various acoustic and prosodic cues, phonological rules, confidence measures and speaking-style information, to name just a few, for the task of mispronunciation detection. Interested readers may also refer to [13-17] for comprehensive and enjoyable overviews of state-of-the-art methods that have been successfully developed and applied to various mispronunciation detection tasks.

The conceptual simplicity and practical effectiveness of end-to-end (E2E) neural networks have recently prompted considerable research efforts into replacing the conventional ASR architectures with integrated E2E modeling frameworks which learn the acoustic and language models jointly rather than separately. Two representatives of E2E models are the connectionist temporal classification (CTC) and the attention-based model. CTC generally consists of two processing stages: an acoustic model (encoder) followed by a letter (phonetic or word-piece)-level translation model (decoder) that generates an output sequence meanwhile preserving the left-to-right alignment order between the input and output sequences under a Markov assumption [18]. On the other hand, an attention-based model is usually composed of an acoustic level encoder and a language level decoder working in tandem, harnessing the power of an attention mechanism to perform soft-alignment (or construct associations) between the hidden acoustic embeddings and the recognized output symbols [19]. By combining the strengths of both CTC and the attention-based model, it is anticipated that the resulting composite model (a.k.a. hybrid CTC-Attention approach) can utilize CTC to assist the attention-based model to compensate for the misalignment problem and improve the speed of the decoding process [20, 21].

More recently, the E2E-based modeling paradigm instantiated with CTC has also been introduced to MD to with promising results, in comparison to the GOP-based method that builds on the hybrid DNN-HMM acoustic model [22]. The training objective of CTC was the same as those used for ASR, while the output of CTC (viz. speech recognition results), in relation to the corresponding text prompt, was directly employed to judge the correctness of phone-level pronunciations and meanwhile provide diagnostic feedback. Motivated by the aforementioned observations, in this paper we first present a novel use of hybrid CTC-Attention approach to the MD task. As such, we can take advantage of the strengths of both CTC and the attention-based model, meanwhile getting around the need for strict phone-level forced-alignment of waveform signals in relation to refence text prompts. Second, we perform input augmentation with text prompt information to make the resulting E2E-based model more tailored for MD. On the other hand, we adopt two MD decision methods to better cooperate with the proposed framework: 1) decision-making based on a recognition confidence measure or 2) simply based on speech recognition results.

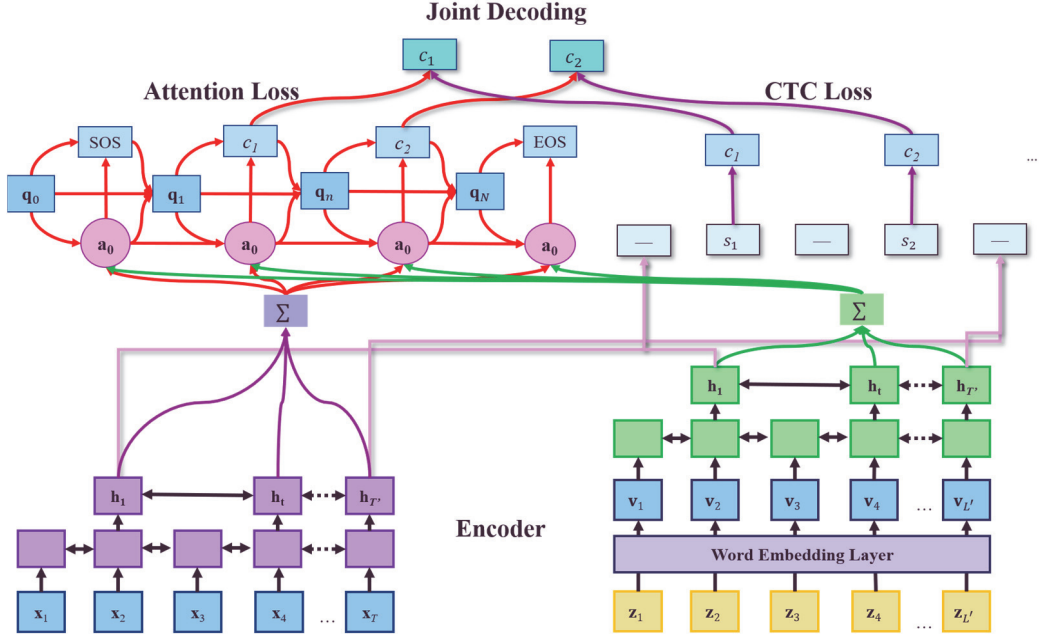

Figure 1: *A schematic depiction of our proposed E2E-based MD system.*

The rest of this paper is organized as follows. We introduce the general procedure of the E2E-based approach to MD in Section 2. Section 3 elucidates the way we perform text-prompt augmentation for use in the MD process. After that, the experimental settings and extensive sets of MD experiments are presented in Sections 4 and 5, respectively. Finally, Section 6 concludes the paper and provides future directions of research.

## 2. End-to-End Approach to MD

### 2.1. CTC Modeling Component

The formulation of CTC directly follows from the Bayes decision rule, viz. finding the output symbol (letter or word-piece) sequence $C = c_1, c_2, \ldots c_L$ that has the maximum posterior probability given the frame-wise input acoustic vector sequence $\mathbf{X} = \mathbf{x}_1, \mathbf{x}_2, \ldots, \mathbf{x}_T$, which can be further factorized as follows:

$$P_{\text{CTC}}(C|\mathbf{X}) \approx \sum_S P(C|S) P(S|\mathbf{X}) \qquad (1)$$

where $S = s_1, s_2, \ldots, s_T$ is a frame-wise, latent symbol sequence, while $P(S|\mathbf{X})$ and $P(C|S)$ are referred to as the acoustic model (or encoder) and translation model (or decoder) of CTC, respectively [18]. The acoustic model $P(S|\mathbf{X})$ can be instantiated with deep neural networks such as bi-directional long short-term memory (BLSTM), Transformer, and among others. Note also here that the logit $\mathbf{h}_t$ corresponding to each frame-wise latent symbol $s_t$ can be regarded as a distilled embedding vector that represents the acoustic characteristics of the input $\mathbf{X}$ at time $t$. On a separate front, the translation model can be embodied with letter (or word-piece) based language model with the first-order Markov assumption.

### 2.2. Attention-based Modeling Component

The attention-based model directly estimates the posterior probability of the output symbol sequence given the input acoustic vector sequence, without making any independence assumption on the output symbol sequence:

$$P_{\text{ATT}}(C|\mathbf{X}) = \prod_{l=1}^{L} P(c_l|\mathbf{X}, c_{1:l-1}) \qquad (2)$$

where $P(c_l|\mathbf{X}, c_{1:l-1})$ can be estimated with an encoder working jointly with a decoder. The former first encodes the acoustic characteristics of an input acoustic vector sequence $\mathbf{X}$ into a latent embedding sequence: $\mathbf{h}_1, \mathbf{h}_2, \ldots, \mathbf{h}_T = \text{Encoder}(\mathbf{X})$. The latter then computes the probability of $c_l$ at each output position by conditioning on the previous decoded symbol sequence $c_{1:l-1}$, the current hidden state representation $\mathbf{q}_l$, as well as the attended vector $\mathbf{r}_l$ derived from the attention mechanism:

$$e_{lt} = \text{Attention}(\mathbf{q}_{l-1}, \mathbf{h}_t, a_{l-1}) \qquad (3)$$

$$a_{lt} = \frac{\exp(\gamma e_{lt})}{\sum_l \exp(\gamma e_{lt})} \qquad (4)$$

$$\mathbf{r}_l = \sum_{t=1}^{T} a_{lt} \mathbf{h}_t \qquad (5)$$

$$P(c_l|\mathbf{X}, c_{1:l-1}) = \text{Decoder}(\mathbf{r}_l, \mathbf{q}_l, c_{l-1}) \qquad (6)$$

where $a_{lt}$ is the normalized attention weight of $e_{lt}$, obtained via the softmax function, and $\gamma$ is a sharpening factor that modulates the distribution of the attention weights. One thing to note is that both CTC and the attention-based model can employ the same model formulation for their encoders.

### 2.3. Hybrid CTC-Attention Approach

Since the attention-based model has the disadvantages of non-monotonous left-to-right alignment and slow convergence, whereas CTC has to use an additional well-trained translation (language) model to achieve better results, there is good reason to integrate them together [20, 21], especially leveraging CTC to help constrain the left-to-right alignment order between the input and output sequences when decoding a possible output sequence $C$:

$$P_{\text{CTC-ATT}}(C|\mathbf{X}) = \lambda P_{\text{CTC}}(C|\mathbf{X}) + (1-\lambda) P_{\text{ATT}}(C|\mathbf{X}) \quad (7)$$

where $\lambda$ is introduced to specify the relative importance of $\text{CTC}(C|\mathbf{X})$ and $P_{\text{ATT}}(C|\mathbf{X})$.

## 3. Text-Prompt Augmentation for E2E-based MD Systems

Our architecture proposed in this section is inspired by [23], which made use of two separate encoders in the ASR architecture: one for acoustic inputs and another for text-based inputs. Different from [23], we use the text prompt of a test utterance for text-based augmentation. This way, inputs to the text-prompt encoder are the associated phone-level-reference text prompts of either training or test utterances, rather than arbitrary synthetic texts. With the novel use of text prompts, we aim to make the resulting hybrid CTC-Attention model more tailored for the MD task. More specifically, if mispronounced training utterances are available in the training phase, it is anticipated that an augmented encoding of the reference text prompt can serve as a clue that makes the trained MD model more concentrated on predicting (detecting) the mispronounced phones. Figure 1 shows a schematic depiction of our proposed E2E-based MD system.

### 3.1. Augmentation of Text Prompts

As depicted in Figure 1, the MD system is augmented with the phone-level symbol sequence $Z = z_1, z_2, \ldots, z_L$ that corresponds to the reference text prompt of a training (or test) utterance, where $z_i$ denotes the symbolic representation of an arbitrary phone $i$. Two disparate instantiations of the phone-level symbol sequence are investigated [23], as highlighted in Table 1:

1. Phone Stream (PS): The text-prompt encoder simply takes the phone-level symbol sequence of a text prompt as the input. Note that, in the training and test phases, it is expected that the output sequence of the network should be the phone-level sequence that a speaker has actually uttered.
2. Repeated Phone Stream (RPS): The phone-level symbol sequence of a text prompt has proper repetitions of each constituent phone, which reflects the relative durations of phones (the corresponding statistics can be obtained with a forced-alignment process) in the training dataset.

Note here that in our implementation, the phone-level symbol sequence $Z = z_1, z_2, \ldots, z_L$ is first transformed to its corresponding vector sequence $\mathbf{v}_1, \mathbf{v}_2, \ldots, \mathbf{v}_L$, before it is fed into the text-prompt encoder, as illustrated in the bottom-right part of Figure 1.

### 3.2. Use of Confidence Measures

In order to compare our method with the conventional GOP-based method, we design a confidence measure for use in our MD system. To this end, we alternatively constrain the hybrid CTC-Attention model, at each time stamp, can only output a phone-level symbol that is involved in the text prompt. As such, for a phone at a given position of the text prompt, we can readily obtain its posterior probability $P(z_l|\mathbf{X})$, which is then taken as an input to a decision function for calculating its confidence measure:

$$D(P(z_l|\mathbf{X})) = \frac{1}{1 + \exp(P(z_l|\mathbf{X}))} \quad (8)$$

$$\mathbb{I}(D(P(z_l|\mathbf{X}))) = \begin{cases} 1 \text{ if } D(P(z_l|\mathbf{X})) \geq \tau \\ 0 \text{ otherwise} \end{cases} \quad (9)$$

Table 1: *Examples of symbol sequences of text prompts under different generation schemes*.

| Token | Example Sequence |
|---|---|
| PhoneStream (PS) | b a1 |
| Rep-PhoneStream (Rep-PS) | b b a1 a1 a1 a1 a1 |

Table 2: *Statistics of the speech corpus used in the mispronunciation detection experiments.*

| | | Duration (hours) | # Spks | # Phn | # Errs. |
|---|---|---|---|---|---|
| Train | L1 | 6.68 | 44 | 73,074 | NA |
| | L2 | 15.79 | 63 | 118,754 | 26,434 |
| Dev | L1 | 1.40 | 10 | 14,216 | NA |
| | L2 | 1.46 | 6 | 11,214 | 2,699 |
| Test | L1 | 3.20 | 26 | 32,568 | NA |
| | L2 | 7.49 | 44 | 55,190 | 14,247 |

When the output of $\mathbb{I}(D(P(z_l|\mathbf{x})))$ is equal to 0, it indicates that $z_l$ has been mispronounced. The thresholding parameter is determined based on the receiver operating characteristic (ROC) curve using the development dataset.

### 3.3. Use of Speech Recognition Results

Apart from the confidence measure discussed in Section 3.2, we can simply recast CAPT as an ASR problem, using the hybrid CTC-Attention model depicted in Figure 1 [22, 24, 25]. In other words, the output phone-level symbol of the hybrid CTC-Attention model at each time stamp is first aligned with a phone-level symbol at some position of the text prompt with the shortest edit distance criterion [26]. If any pair of such alignment contains phone-level labels that are not identical, it signifies that a phone-level mispronunciation occurs. As a side note, in this study, we focus exclusively on detecting mispronunciations caused by substitution and deletion errors, since the number of insertion errors are relatively very few in the speech corpus to be introduced in Section 4.1.

## 4. Experimental Setup

### 4.1. Experimental Corpus

The dataset employed in this study is a Mandarin annotated spoken (MAS) corpus compiled by the Center of Learning Technology for Chinese, National Taiwan Normal University, between 2012 and 2014 [27]. The corpus was split into three subsets: training set, development set and test set. All these subsets are composed of speech utterances (containing one to several syllables) pronounced by native speakers (L1) and L2 learners. Each utterance of an L2 learner may contain mispronunciations, which were carefully annotated by at most four human assessors with a majority vote. Table 1 briefly highlights the statistics of the speech corpus. Our MD system with the hybrid CTC-Attention model was built on the ESPnet toolkit [28], while the GOP-based MD system with the DNN-HMM model was built on the Kaldi toolkit [29].

### 4.2. Evaluation Metrics

The default evaluation metric employed in this paper is the F1-score, which is a harmonic mean of precision and recall, defined as:

$$F1 = \frac{2 * Precision * Recall}{Precision + Recall} \quad (11)$$

$$Precison = \frac{C_{D \cap H}}{C_D} \quad (12)$$

$$Recall = \frac{C_{D \cap H}}{C_H} \quad (13)$$

where $C_D$ is the total number of phone segments in the training set that were identified as being mispronounced by the current mispronunciation detection module, $C_H$ is the total number of phone segments in training set that were identified as being mispronounced by the majority vote of human assessors, and $C_{D \cap H}$ were is the total number of phone segments in the training set that are identified as being mispronounced simultaneously by both the current mispronunciation detection module and the majority vote of human assessors.

## 5. Experimental Results and Discussion

### 5.1. E2E MD with Confidence Measures and Recognition Results

At the outset, we evaluate the performance of two variants of our baseline MD system built on top of a hybrid CTC-Attention model (denoted by CTC-ATT), in comparison to the celebrated GOP-based method, as well as its extension that leverages maximum F-1 criterion (MFC) [30]. The first variant of our baseline MD system performs MD with a confidence measure calculated based on the output of a constrained hybrid CTC-Attention model (*cf.* Section 3.2), denoted by CONF hereafter. The second one that recast MD as an ASR task (*cf.* Section 3.3), denoted by SR hereafter. Both these two variants do not resort to the text-prompt encoder (*cf.* Section 3.1). As can be seen from Table 3, CTC-ATT(CONF) performs on par with GOP, but is inferior to GOP+MFC that further employs MFC to train its component DNN-HMM model. However, CTC-ATT(SR) outperforms both GOP and GOP+MFC by a big margin. This also reveals that E2E-based modeling, such CTC-ATT, holds excellent promise for the MD task. In addition, CTC-ATT(SR) shows its superiority over CTC(SR) (viz. performing MD merely with CTC) in terms of the precision performance.

In the second set of experiments, we turn to investigating whether it is beneficial to relieve the restriction of determining the correctness of a phone-level pronunciation strictly based on the one-best phone hypothesis at the associated (aligned) time stamp, generated from CTC-ATT(SR). Nevertheless, it is clearly evident from Table 4 that significant drops of the MD performance occur when we relax our standards to determine the correctness of a phone-level pronunciation by matching it with any one of the *N*-best (*N*=2 to 5) phone hypotheses at the associated time stamp to see if they have the identical phone label.

Table 3: *Performance of various E2E-based MD methods and the existing GOP-based methods.*

|  | Recall | Precision | F1 |
|---|---|---|---|
| GOP | 0.518 | 0.635 | 0.570 |
| GOP+MFC | 0.695 | 0.613 | 0.652 |
| CTC-ATT(CONF) | 0.689 | 0.509 | 0.586 |
| CTC-ATT(SR) | 0.708 | **0.679** | **0.692** |
| CTC(SR) | 0.706 | 0.656 | 0.680 |

Table 4: *Performance of CTC-ATT(SR) when using the N-best results for the MD task.*

|  | Recall | Precision | F1 |
|---|---|---|---|
| CTC-ATT(SR) | **0.708** | 0.679 | **0.692** |
| +2-best | 0.534 | 0.743 | 0.621 |
| +3-best | 0.445 | 0.777 | 0.566 |
| +4-best | 0.385 | 0.794 | 0.518 |
| +5-best | 0.346 | 0.803 | 0.483 |

Table 5: *Performance of CTC-ATT(SR) when two variants of text-prompt augmentation are exploited, respectively.*

|  | Recall | Precision | F1 |
|---|---|---|---|
| CTC-ATT(SR) | 0.708 | 0.679 | 0.692 |
| +PS | **0.718** | **0.684** | **0.702** |
| +RPS | **0.718** | **0.684** | 0.701 |

### 5.2. Augmentation of Text Prompts

In the last set of experiments, we assess the effectiveness of augmenting the associated text prompt of a given utterance for use in our MD system (*cf.* Section 3.1). Two variants of the text-prompt augmentation approach are evaluated here: the phone-stream method (PS) and the repeated phone-stream method (RPS). This approach is inspired from the conventional GOP-based methods that make use of the reference text prompt; for the GOP-based methods, the required forced-alignment information of a given test utterance, which is exploited during the MD process, is obtain by referring to its associated text prompt. The corresponding results of text-prompt augmentation are shown in Table 5. There are two noteworthy points to these results. First, the additional use of text-prompt augmentation can offer a relative F1-score improvement of roughly 1.5%. Second, the two variants of text text-prompt augmentation are quite competitive with each other. We believe that further investigation of more elaborate text-prompt augmentation for the E2E-based MD task would still be desirable.

## 6. Conclusion and Outlook

In this paper, we put forward a novel end-to-end modeling paradigm for mispronunciation detection (MD), for which disparate model structures and variants of text-prompt augmentation have been investigated. Extensive sets of experiments have been conducted to evaluate the effectiveness of the various MD systems stemming from the proposed modeling paradigm; the associated results indeed reveal their performance merits, in comparison to the well-practiced GOP-based method and its extension. In future work, we plan to explore more elaborate text-prompt augmentation approaches and other effective optimization criteria for training the component models of our E2E-based MD system.